\documentstyle[aps,epsfig]{revtex}
\begin{document}

\title{Feedback effects and the self-consistent Thouless criterion \\
of the attractive Hubbard model} 
\author{K.S.D. Beach\cite{byline}, R.J. Gooding}
\address{Department of Physics, Queen's University,\\
Kingston, Ontario, Canada K7L 3N6}
\author{F. Marsiglio}
\address{Department of Physics, University of Alberta,\\
Edmonton, Alberta, Canada T6G 2J1}

\date{\today}
\maketitle

\begin{abstract}
We propose a fully microscopic theory of the anomalous normal state of the 
attractive Hubbard model in the low-density limit that accounts for propagator 
renormalization. Our analytical conclusions, which focus on the thermodynamic 
instabilities contained in the self-consistent equations associated with our
formulation, have been verified by our comprehensive numerical study of the 
same equations. The resulting theory is found to contain no transitions
at non-zero temperatures for all finite lattices, and we have confirmed,
using our numerical studies, that this behaviour persists in the thermodynamic 
limit for low-dimensional systems.
\end{abstract}

The anomalous normal state properties of the high temperature
superconductors have, amongst other things, highlighted the limitations
of present-day many-body physics. In spite of repeated attempts
\cite{kadanoff61,baym61,bickers89} to formulate a controlled
fully microscopic many-body expansion, at present both microscopic and 
semi-phenomenological theories require vindication from Monte Carlo or 
some other ``exact" procedure. 

To serve as further and concrete motivation for the work described below,
consider the paper of Schmitt-Rink, Varma, and Ruckenstein
\cite{schmitt-rink89}. They proposed, through a simple formulation of the 
problem, that in two
dimensions (2d) any electronic effective two-particle interaction 
that was attractive would lead to a destabilization of the Fermi 
surface, and thus to non-Fermi liquid-like physics.
One may consider this result as being based on
the idea that the Fermi surface is destroyed via the
presence of two-particle bound states that ``drain away" all electrons
to the bound state. The result is a superconducting instability 
which, at best, occurs only at zero temperature.
The $T=0$ predictions of this model are complicated: they found a
critical point, corresponding to all electronic densities,
with all electrons paired --- whether or not these paired electrons
were in a phase coherent state was not addressed, although that is
what one would expect.

Many questions and criticisms of this approach have appeared
\cite {randeria}.
In particular, by including multiple scattering in the electron
number equation, Serene \cite{serene89} claimed that the physics of 
Ref. \cite{schmitt-rink89} was eviscerated, and that Fermi liquid
behaviour was in fact robust. Actually, as we show below, at the
multiple-scattering level of formulation the physics of 
Ref. \cite{schmitt-rink89} survives such an improvement.

Another criticism of the result of Schmitt-Rink, Varma, and Ruckenstein
\cite {schmitt-rink89} is that Luttinger's theorem is not satisfied ---
this immediately suggests that renormalization of the single-particle
propagators is important. As we show below, this is indeed the case, and,
in fact, the physics of Ref. \cite {schmitt-rink89} is dramatically changed 
when such renormalization, or self-consistency, is included.
The problem with such a proposal is that the renormalization of
these propagators is not a simple matter, and in this paper
we address the question of the ``appropriate'' self-consistent theory,
and apply our considerations to the attractive Hubbard model (AHM).

The impact of self-consistency on the results of Ref. \cite {schmitt-rink89}
is related to the more general question of the study of feedback
effects on the Thouless criterion. The Thouless criterion arises from the
determination, with decreasing temperatures in the normal state, of the 
superconducting instability \cite{thouless60}. Such a formulation 
traditionally uses unrenormalized electron propagators --- see 
\cite {ambegaokar69} for a particularly clear discussion of the
physics that arises from such an approach. An important advantage of the 
Thouless approach is that it focuses on the non-superconducting state, so the 
possibility of non-Fermi liquid physics in the anomalous normal state can 
be studied simultaneously with the question of the superconducting 
instability. Thus, the issue of the self-consistent Thouless criterion
can be addressed by repeating the analysis of, say, Ref. 
\cite {ambegaokar69}, using renormalized propagators. 
This provides feedback, {\em i.e.} the
single-electron propagators will ``know" about the pairing tendency
of the system as the temperature is lowered, and this in turn will
influence, for better or for worse, that pairing tendency.

What self-consistent formulation is appropriate?  One suggestion comes 
from Levin and coworkers \cite{janko97,levin...}, based on earlier 
work \cite{kadanoff61,schmid70,marcelja70,patton71}, in which only one
specific combination of renormalized propagators (in the equations for
both the pair susceptibility and the self energy) was proposed.
Fresard, {\it et al.} \cite {fresard92}, and then Haussmann \cite {haussmann} 
(and several subsequent authors),
examined the situation that arises when all propagators
are renormalized. Another altogether different approach was adopted by Vilk 
and Tremblay \cite{vilk97}, who used sum rules, symmetry, and other 
considerations, to construct a theory in 2d which required
that renormalized propagators {\em not} be used. Our work
is different from all of these attempts.

For electron densities near half-filling there is no justification for
omitting vertex corrections. However, the approach we adopt is designed 
specifically for low electron densities, for which we have demonstrated
that vertex corrections are negligible. We find (i) no superconducting
transition at non-zero temperatures for all finite lattices, curing
a difficulty that is present with all mean-field theories;
(ii) a self-consistent solution at $T=0$ that corresponds to a superconducting 
phase; and (iii) numerical solutions of the resulting self-consistent
equations confirm our analytical results and demonstrate that in low
dimensions our conclusions survive in the thermodynamic limit. Further,
based on our numerical solution we predict the low-temperature form of the
pair susceptibility.

The above-mentioned papers examined a variety of different Hamiltonians.
For concreteness, we used the attractive Hubbard model (AHM) on
a d-dimensional hypercubic lattice, defined by
\begin{equation}
\label{eq:ahm ham}
H ~=~-t~\sum_{\langle{\bf m, m}^\prime\rangle,\sigma}
c^\dagger_{{\bf m},\sigma} c_{{\bf m}^\prime,\sigma} 
~-~|U|~\sum_{\bf m} n_{{\bf m},\uparrow} n_{{\bf m},\downarrow}
\end{equation}
where the terms in this Hamiltonian have their usual meaning.

When the electronic density is either close to zero 
($\langle \hat n_{\bf m} \rangle \approx 0$), or 
complete filling  ($\langle \hat n_{\bf m} \rangle \approx 2$),
the effective interaction is well described by the T-matrix
approximation \cite{fetter71}. The T-matrix approximation
corresponds to a self-energy given by (ignoring, for this
publication, the Hartree term)
\begin{equation}
\label{eq:se}
\Sigma({\bf k},i\omega_n)=-\frac{U^2}{N\beta}
\sum_{{\bf q},i\Omega_\ell}~
{\cal{G}}_2({\bf q},i\Omega_\ell)
~G_{\alpha}({\bf -k+q},i\Omega_\ell-i\omega_n)~~,
\end{equation}
where  ${\cal{G}}_2$ is the pair propagator, and $G_{\alpha}$ 
can be either $G_0$ (the non-interacting) or $G$
(the renormalized) single-particle Green's function. 
The pair propagator is given by
\begin{equation}
\label{eq:Tmatrix}
{\cal{G}}_2({\bf q},i\Omega_\ell)~=
~\frac{\chi({\bf q},i\Omega_\ell)}
{1~-~|U|~\chi({\bf q},i\Omega_\ell)}~~,
\end{equation}
where the pair susceptibility, $\chi$, is a convolution of two
single-particle propagators given by
\begin{equation}
\label{eq:chi}
\chi({\bf q},i\Omega_\ell)=\frac{1}{\beta N}\sum_{{\bf k},i\omega_n}
G_{\alpha^{\prime}}({\bf k},i\omega_n)
G_{\alpha^{\prime \prime}}({\bf q-k},i\Omega_\ell-i\omega_n)~~,
\end{equation}
with $G_{\alpha^\prime}$ and $G_{\alpha^{\prime \prime}}$ 
either $G_0$ or $G$.
In these equations, $\beta$ is the inverse temperature ($k_B=1$), $N$ is 
the number of lattice sites, and $\omega_n$ ($\Omega_\ell$) 
are Fermi (Bose) 
Matsubara frequencies. Finally, Dyson's equation gives another relation
between the self-energy and the renormalized Green's function, $G$. 

As outlined in Ref. \cite {ambegaokar69},
the Thouless criterion, signifying the instability of the normal
state, corresponds to poles of the zero-centre-of-mass momentum retarded, 
real-time, pair propagator moving off the real axis into the upper half of the
complex plane. From Eq.~(\ref{eq:Tmatrix}) this simplifies to the condition
(defining $\chi \equiv \chi^\prime + \chi^{\prime\prime}$)
\begin{equation}
\label{eq:Th Crit}
\chi^\prime({\bf q}=0,i\Omega_\ell=0)~=~1/|U|~~,
\end{equation}
(noting that $\chi^{\prime\prime}({\bf q}=0,i\Omega_\ell=0)~=~0$).

For future reference, let us make clear the instability that is
found when propagator renormalization is ignored. Traditionally
\cite {ambegaokar69}, one evaluates Eq.~(\ref{eq:chi}) using
bare propagators. The divergence of Eq.~(\ref{eq:Tmatrix}) then
leads to an instability at a temperature $T_c$ which is identical
to that predicted by BCS theory ({\it e.g.,} see Ref. \cite {ambegaokar69}). 
Below we discuss some of the effects of feedback on this scenario.

As further information on the above no-feedback theory,
note that researchers have begun to use the appearance of the pseudogap
as an important test for the presence of strong correlations in
model systems, and this requires that one obtain the dynamics, and
not the static thermodynamic quantities, for the proposed model. However, 
to obtain the dynamics from even the simple non-self-consistent T-matrix
theory involves some hard-to-control extrapolations to the real
frequency axis. Recently, we
formulated a partial fraction decomposition method,
that is described elsewhere \cite{houstonbeach},
and which can be evaluated with a precision (with respect to pole 
locations and residues of the self energy) to a relative accuracy of 
$10^{-80}$. The resulting density of states 
shows a well-defined pseudogap at low temperatures \cite{houstonbeach}, 
very much reminiscent of experiment \cite{timusk99}. Further,
our results for the single electron spectral functions derived from the 
non-self-consistent T matrix are remarkably similar in character to the 
quantum Monte Carlo/maximum entropy method spectral functions and density of
states determined by the Sherbrooke collaboration
\cite {allen99} (although their calculations are for a much higher
electron filling). Thus, from a comparison to either experiment
or ``exact" Monte Carlo data, it appears that 
a T-matrix theory with unrenormalized propagators is a reasonable
formalism with which to examine this problem. However, as discussed
at the start of this paper (with regards to Ref. \cite {schmitt-rink89}), 
and as we now delineate below, this agreement is fortuitous --- 
the predictions for the (static) thermodynamics for such
a theory are in error, and thus the dynamics produced by such
a formulation are of no value.

To see how the thermodynamics predicted by 
Refs.~\cite {schmitt-rink89,serene89} make clear
the failings of such a theory, consider the following: If one
chooses a chemical potential that lies in the band ($|\mu|<W/2$,
where $W$ is the single-particle non-interacting bandwidth), and then
lowers the temperature, using the original Schmitt-Rink {\em et al.} 
\cite{schmitt-rink89} formulation applied to the 2d AHM, we have shown
that the electron density actually diverges as the instability
is approached (for constant a chemical potential). Use of the
correct number equation, as follows from the number-conserving theory of 
Serene \cite {serene89}, leads to a number density of one at the instability 
temperature (see below). Then, this leads to the unphysical result
that for any fixed electron density less than one, 
the chemical potential is always driven to half the bound state
value ({\em viz.} below the band),
indicative of the elimination/suppression of the Fermi surface.
Further problems abound ---   while the electron density
approaches unity at the instability, the two-particle correlation function 
$\langle \hat n_{\bf m,\uparrow } \hat n_{\bf m, \downarrow }\rangle$
diverges --- that is, this formulation has difficulties at ``the two-particle 
level" \cite {vilk97}. This is clearly incorrect, since for the AHM it 
should be bounded between $\langle \hat n_{\bf m} \rangle^2/4$ and 
$\langle \hat n_{\bf m} \rangle/2$.

\begin{figure}
\label{fig:muT plane}
\begin{center}
\epsfig{file=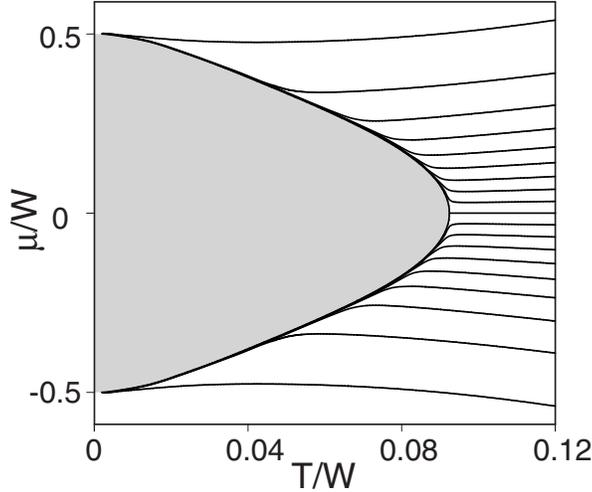,scale=0.95}
\caption{Lines of constant density
are plotted in the $\mu$-$T$ plane (each quantity shown in
units of the bandwidth $W$) for a $14 \times 14$
lattice. The contours shown here (from bottom to top)
correspond to electron densities $\langle n \rangle
=0.1,0.2,0.3,\dots,1.8,1.9$
and are calculated for $|U|/t=4$.  The Thouless criterion line
forms the boundary to the shaded superconducting region.
Notice that all the density contours, except the $\langle n \rangle=1$
line, are deflected away from this area, and never cross into
this area, as they proceed to lower temperatures.}
\end{center}
\end{figure}

We now demonstrate the above-mentioned behaviour analytically:
We note that Eq.~(\ref{eq:Th Crit}) indicates
that near the Thouless curve ($T_c$) the self-energy (Eq.~(\ref{eq:se})) is
dominated by the ${\bf q}=0,~\Omega_\ell=0$ term.
Thus for all finite lattices one finds
\begin{equation}
\label{eq:se near Tc}
\Sigma({\bf k},i\omega_n)\sim-\frac{U^2}{N\beta}
\frac{\chi(0,0)}{1-|U|~\chi(0,0)}~G_0({\bf -k},-i\omega_n)~~.
\end{equation}
Note the electron density equation is
\begin{equation}
\label{eq:n eq 1}
\langle \hat n_{{\bf m}}\rangle
=1+{2 \over N\beta}{\cal{R}}e\sum_{{\bf k},i\omega_n}
(i\omega_n-\xi_k-\Sigma({\bf k},i\omega_n))^{-1}~~,
\end{equation}
where $\xi_{\bf k}$ is the band energy relative to the chemical
potential, and that at the Thouless temperature the self-energy diverges 
(Eq.~\ref{eq:se near Tc}),
and thus $\langle \hat n_{\bf m} \rangle = 1$, as stated above
\cite{higherd}. Several
example flow lines are shown in a $(\mu,T)$ plot for a 2d square
lattice in Fig.~\ref{fig:muT plane},
in which it is seen that they approach the two-electron (hole) bound 
state energy below (above) the band as $T\rightarrow 0$. This results
in a system of bound electrons (holes), and the Fermi surface (and hence
the normal state Fermi liquid) is destroyed.

To remedy these deficiencies within a fully microscopic diagrammatic
theory, renormalized propagators are required.
Unfortunately, there are a variety of different
self-consistent T-matrix theories. These different theories are
based on the choice of a non-interacting $G_0$ or a fully
renormalized $G$ in the $G_{\alpha}$'s that appear in 
Eqs.~(\ref{eq:se},\ref{eq:chi}). Based on the consideration of
Mar\v celja \cite {marcelja70} and Patton \cite {patton71}, Levin
and coworkers \cite{janko97,levin...} have argued that the
$G_{\alpha}$ in Eq.~(\ref{eq:se}) must be the
non-interacting Green's function (to avoid states in the superconducting
gap, which are absent in the weak coupling limit). While we agree with
this ``non-derivable'' approach, we disagree with the claim that the
appropriate combination of Green's functions in the pair susceptibility
should be asymmetric ($\chi \sim G~G_0$). Firstly, these theories
do not produce the full two-particle Green's function, but rather just
the pair propagator. One cannot say that a pair propagator is
conserving, but one could ask: Is the two-particle Green's function
from which the pair propagator is formed conserving? It follows from
Ref. \cite{baym61} that a symmetric convolution ($\sim G~G$) leads to 
two-particle quantities that are conserving, and thus it is
not required, nor necessarily wise, to use the asymmetric form ($\sim
G~G_0$) \cite {janko97,levin...} for the pair propagator.

Based on these (and other \cite {franklongTM}) considerations,
we examine the self-consistent theory given by Dyson's equation
with $G_\alpha = G_0$ and $G_{\alpha^{\prime}} = G_{\alpha^{\prime \prime}}
= G$ in Eqs. (\ref{eq:se},\ref{eq:chi}). 
Following our earlier arguments, for non-zero temperatures and a finite 
lattice we can write the ansatz \cite{levin2} near the superconducting 
instability
\begin{equation}
\label{eq:sc ansatz1}
\Sigma({\bf k},i\omega_n)~=~\frac{\Delta^2}{i\omega_n + \xi_{\bf k}}~~,
\end{equation}
\begin{equation}
\label{eq:sc ansatz2}
\Delta^2=\frac{U^2}{N\beta}
\frac{\chi({\bf 0},0)}{(1-|U|\chi({\bf 0},0))}~~.
\end{equation}
From Eq.~(\ref{eq:Th Crit}) it follows that a $T_c > 0$ instability is 
indicated by $\Delta \rightarrow \infty$. However, evaluating
$\chi({\bf 0},0)$ using Eq.~(\ref{eq:sc ansatz1}) in Eq.~(\ref{eq:chi})
leads to
\begin{equation}
\label{eq:sc chi eqn}
\chi({\bf 0},0)~=~\frac{1}{N \beta}~\sum_{{\bf k},\omega_n}~
\frac
{\omega_n^2~+~\xi_{\bf k}^2}
{(\omega_n^2~+~\xi_{\bf k}^2~+~\Delta^2)^2}~~,
\end{equation}
and shows in fact, that $\chi({\bf 0},0)$ goes to zero in the limit
of $\Delta \rightarrow \infty$. Thus, there is no self-consistent
solution at non-zero temperatures for any finite lattice --- this
cures an obvious problem that exists for mean-field (non-self-consistent
T-matrix) theories \cite {3d}. 

The only self-consistent solution of these equations occurs at $T=0$.
That is, by solving $1 = |U|\chi({\bf 0},0)$ in the $T\rightarrow 0$ limit 
we obtain a modified gap equation given by
\begin{equation}
\label{eq:t=0 sc soln}
1=\frac{|U|}{2N}~\sum_{\bf k}~\Big(
\frac{1}{\sqrt{\Delta^2~+~\xi_{\bf k}^2}}
~-~\frac{1}{2}\frac{\Delta^2}{(\Delta^2~+~\xi_{\bf k}^2)^{3/2}}
\Big)~~.
\end{equation}
One finds that in the small $|U|/t$ limit,
$\Delta \sim \exp(-4W/|U|)$;
this shows, in contrast to what was claimed in Ref. \cite{janko97,levin...},
that when both propagators in the pair susceptibility
are renormalized one still recovers the weak coupling BCS
limit \cite{prefactor}. Indeed, the form of the density of states at
$T=0$,
\begin{equation}
\label{eq:T=0 DOS}
{\cal N}(\omega)\sim{\cal
N}^0(0)\frac{|\omega|}{\sqrt{\omega^2-\Delta^2}}
\Theta(\omega^2-\Delta^2)~~,
\end{equation}
is that of weak coupling BCS theory, and shows a fully developed gap.

In order to further analyze the predictions of our theory, we have
conducted a comprehensive numerical study of the resulting
self-consistent equations, and have solved these equations down to 
the lowest temperatures that we can access (typically, these
temperatures are 1/100th of the mean-field BCS temperature). 
Further, we have obtained convergence of all static quantities to 
better than 1 part in $10^5$ (at low temperatures this requires 
that our Matsubara frequency sums be performed with a frequency 
cutoff of $4W$, and we have thus used this same cutoff at all temperatures).
An example set of data is shown in Figure 2
for $d=1$; lattices from $8\times 1$ to $128\times 1$ were studied,
and it was found that for $L\times 1$ lattices, for $L \geq 64$ no
changes with lattice size were encountered. Thus, these data represent 
the quantitative predictions of our renormalized propagator theory for 
one dimension in the thermodynamic limit (with the shown parameter
set). As is clear from this Figure, the $T\rightarrow 0$ limit of 
$\chi({\bf 0},0)$ is $1/|U|$ (in units of $1/t$); however,
$\chi({\bf 0},0)$ remains less than $1/|U|$, and thus no instabilities
are encountered, for all nonzero temperatures. Clearly, as is predicted
by our analytic theory, there is only a zero temperature transition to a 
superconducting state. 

\begin{figure}
\label{fig:chiT}
\begin{center}
\epsfig{file=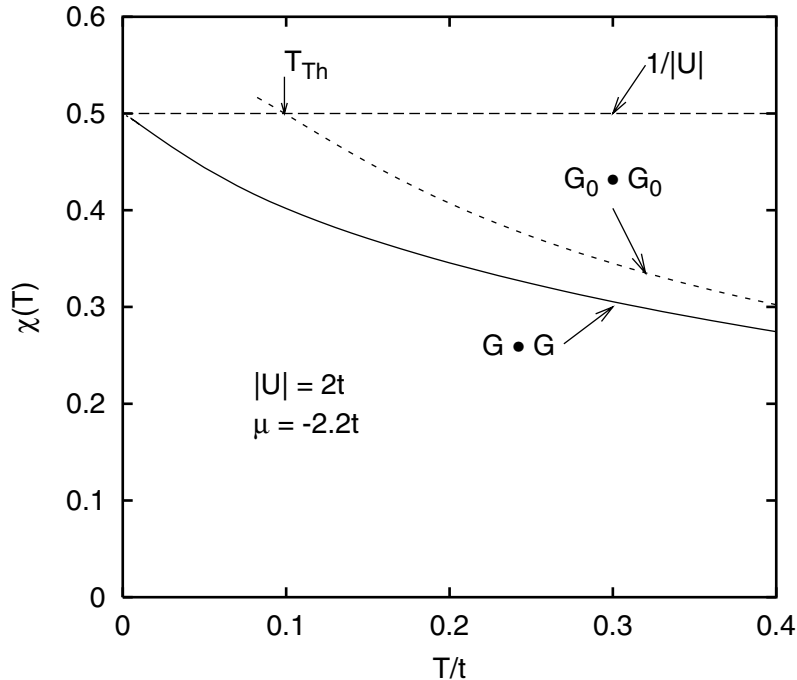}
\caption{$\chi(T) \equiv \chi({\bf q}= 0,i\Omega_\ell = 0)$ 
$vs.$  temperature, T, both in units of $1/t$, for the parameters 
indicated. The Thouless temperature, 
T$_{\rm Th}$, in the (unrenormalized) non-self-consistent T-matrix theory 
(G$_0 \cdot $G$_0$) is indicated with 
an arrow; by including renormalization, this temperature is driven to zero 
(curve indicated with G $\cdot$ G). We have plotted
results for $32\times~1,~64\times~1$ and 
$128\times 1$ lattices; the results are essentially
indistinguishable (and are in fact plotted on top of one another), 
and all extrapolate to $1/|U|=0.5/t$ at $T/t = 0$, to within an
accuracy of better than 0.01\%.}
\end{center}
\end{figure}

Similar data and similar trends are found in two dimensions. However,
if we impose the same strict convergence criterion that we did for
our 1d results (for systems in the thermodynamic limit) we find
that we cannot faithfully extrapolate to the thermodynamic limit in 2d.
For example, using a constant number of Matsubara frequencies leads to 
converged solutions which are unstable with respect to the inclusion
of more frequencies, while using a fixed energy range
(as we did for 1d) leads to limitations in our computing power. 
(At present we are exploring different strategies to overcome this
restriction.) Thus, the behaviour of this formulation in the
thermodynamic limit in two dimensions is an open problem, and whether or
not these numerics confirm the interesting Kosterlitz-Thouless
prediction of Ref. \cite {engel} is unknown.

In conclusion, we have formulated a self-consistent theory with renormalized 
propagators of the AHM. We employ a symmetric pair propopagator formed with 
renormalized single-particle Green's functions $G$, but ``close" the 
self-energy with an unrenormalized $G_0$. We have found a superconducting 
solution at $T=0$, but for all finite lattices we have proven
analytically and confirmed numerically that no non-zero temperature
instabilities exist. The resulting zero-temperature superconductor has a 
conventional density of states in the weak coupling limit. Although it is 
unclear if such a theory accounts for critical effects at $T=0$, it is 
clear that this formulation represents a very useful starting point of a fully 
microscopic theory for the evaluation of, {\it e.g.}, the spectral functions 
at non-zero temperatures (albeit for low electronic densities).

We wish to thank Andr\'e-Marie Tremblay, Gene Bickers, Claude 
Bourbonnais, Boldiszar Janko, and Martin Letz for helpful discussions. 
This work was supported in part by the NSERC of Canada.


\begin{thebibliography}{0}
\bibitem{prefactor} We note that the prefactor of the gap for the 
small $|U|/t$ limit changes by the constant $\sqrt{2}$
when one includes the last term in Eq.~(\protect\ref{eq:t=0 sc soln}).
However, given the inability of BCS, in the weak coupling limit at $T=0$ in
reduced dimensionality transitions, to predict the correct quantitative form 
for the prefactor, this is not a concern --- see, F. Marsiglio, Phys. Rev. B
{\bf 55}, 575 (1997).
\end{thebibliography}
\end{document}